\documentclass[twocolumn,english,showpacs,superscriptaddress,floatfix,nofootinbib]{revtex4}
\usepackage[T1]{fontenc}
\usepackage[latin9]{inputenc}
\usepackage{float}
\usepackage{amsmath}
\usepackage{graphicx}
\usepackage{amssymb}
\usepackage{esint}

\makeatletter
\@ifundefined{textcolor}{}
{%
 \definecolor{BLACK}{gray}{0}
 \definecolor{WHITE}{gray}{1}
 \definecolor{RED}{rgb}{1,0,0}
 \definecolor{GREEN}{rgb}{0,1,0}
 \definecolor{BLUE}{rgb}{0,0,1}
 \definecolor{CYAN}{cmyk}{1,0,0,0}
 \definecolor{MAGENTA}{cmyk}{0,1,0,0}
 \definecolor{YELLOW}{cmyk}{0,0,1,0}
 }

\usepackage{epsfig}\usepackage{babel}\usepackage{babel}\usepackage{babel}\usepackage{babel}\usepackage{babel}\usepackage{babel}\usepackage{babel}\usepackage{babel}\usepackage{babel}

\makeatletter
\@ifundefined{textcolor}{}
{
 \definecolor{BLACK}{gray}{0}
 \definecolor{WHITE}{gray}{1}
 \definecolor{RED}{rgb}{1,0,0}
 \definecolor{GREEN}{rgb}{0,1,0}
 \definecolor{BLUE}{rgb}{0,0,1}
 \definecolor{CYAN}{cmyk}{1,0,0,0}
 \definecolor{MAGENTA}{cmyk}{0,1,0,0}
 \definecolor{YELLOW}{cmyk}{0,0,1,0}
 }
\makeatother
\makeatother
\makeatother
\makeatother
\makeatother

\makeatother

\usepackage{babel}

\makeatother

\usepackage{babel}

\makeatother

\usepackage{babel}

\makeatother

\usepackage{babel}

\begin{document}

\title{Anomalous proximity effect in spin-valve SFF structures }

\author{T.~Yu.~Karminskaya}

\address{Skobeltsyn Institute of Nuclear Physics, Moscow State University,
Moscow, 119992 Russia}

\author{A.~A.~Golubov}

\address{Faculty of Science and Technology and MESA+ Institute of Nanotechnology,
University of Twente, P.O.~Box~217, 7500 AE Enschede, The~Netherlands}

\author{M.~Yu.~Kupriyanov}

\address{Skobeltsyn Institute of Nuclear Physics, Moscow State University,
Moscow, 119992 Russia}
\begin{abstract}
We investigate SFF structures (S-superconductor, F feromagnetic metal)
with noncollinear magnetizations of F films with arbitrary transparency
of FF interface. We show the existence of phase slips both at SF and
FF interfaces which manifest themselves in the anomalous dependence
of the spin-triplet correlations on misorientation angle between magnetization
vectors in the F-layers. We discuss how these effects can be observed
in experiments with Josephson $\pi$-junctions.
\end{abstract}


\maketitle
janaph@gmail.com

Nowadays there is a considerable interest to the structures composed
from superconducting (S) and ferromagnetic (F) layers \cite{GKI}-\cite{bverev}.
The possibility of $\pi$-states in SFS Josephson junctions due to
oscillatory nature of superconducting order parameter induced into
a ferromagnet was predicted theoretically \cite{Buz0}-\cite{Buz2}
and has been convincingly demonstrated by experiments \cite{ryazanov2001}-
\cite{Blamire3}. It was also shown recently that both classical and
quantum circuits \cite{QQ1}-\cite{RSFQ} can be realized using SFS
sandwich technology \cite{ryazanov2001}. A number of new phenomena
were predicted in junctions with more than one magnetically ordered
layer. Particularly interesting are equal-spin triplet superconducting
correlations which can penetrate a ferromagnet on a long-range scale
\cite{trip1}-\cite{Asano}. These states are generated if spin rotation
symmetry is broken and therefore are expected to become most important
when angle $\alpha$ between magnetization vectors of ferromagnetic
layers is close to $\pi/2.$ Long-range triplets were recently realized
experimentally in Josephson junctions in a number of geometries and
material combinations \cite{Keizer}-\cite{AartsLR} and in SFF spin
valves \cite{nowak}-\cite{Fominov}.

It was also predicted that in Josephson junctions with several ferromagnetic
layers it is possible to realize $\pi$-states even in the case when
the F-layers are so thin that order parameter oscillations can not
develop there, but phase slips occur at the SF interfaces with finite
transparency. This effect was predicted in Ref.\cite{GKF2002} for
SFIFS junctions, where two SF-bilayers are decoupled by an insulating
barrier 'I'. In this case phase shifts $\delta\phi$ occur at each
of the SF interfaces and saturate at $\delta\phi=\pi/2$ with the
increase of exchange field. As a result, total phase shift across
the junction equals to $\pi$.

Recently, structures where two F-layers are coupled to a superconductor
(FSF or SFF) attracted much attention since they may serve as superconducting
spin valves, where transition temperature is controlled by angle $\alpha$
between magnetization directions of the F-layers. The SFF structures
with fully transparent interfaces were studied theoretically in \cite{Fominov}
where it was shown that critical temperature $T_{c}$ in such trilayers
can be a nonmonotonic function of the angle $\alpha$.

In this paper we address important issue of the influence of interface
transparency on singlet and triplet correlations in SFF structures
and show that the interface phase slips can lead to a number of new
peculiar phenomena. First, the magnitudes of singlet and long-range
triplet components which are generated in SFF structures with varying
angle $\alpha$ between the F-layer magnetizations, have anomalous
dependence on $\alpha$. Namely, contrary to the previous knowledge
based on analysis of symmetric FSF or SFFS structures, the triplet
component in SFF structures reaches maximum not in the vicinity of
$\alpha=\pi/2$ and can be even zero for this configuration of magnetization
vectors. Second, $\pi$-state in SFFIS Josephson junction can be realized
for parallel orientations of magnetizations in the F-layers as a result
of phase shifts at the interfaces.

\begin{figure}[h]
 \centerline{\includegraphics[scale=0.3]{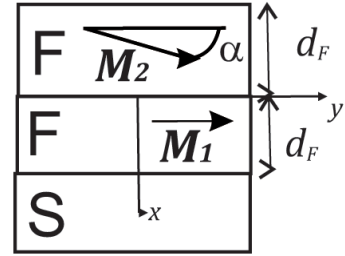}} \caption{ SFF structure}

\end{figure}

To prove the above statements we consider SFF structure presented
in Fig.1. It consists of two identical single domain ferromagnetic
films, which may differ only by a value of exchange energy, $H_{1},$
and $H_{2}$ for lower and upper layer respectively. The magnetization
vector of lower F film is directed along $y$ axis, while in the upper
film it may be deflected by angle $\alpha$ from this direction in
the $yz$ - plane. We will also suppose that the condition of dirty
limit is valid for all the films and that the transparency of SF interface
is small enough providing the opportunity to use linearized Usadel
equation in the form \cite{bverev}, \cite{trip1}. So, for upper
F film we have:

\begin{equation}
\begin{gathered}\xi_{F}^{2}\nabla^{2}f_{0}-\Omega f_{0}-ih_{2}\cos\alpha f_{3}=0,\\
\xi_{F}^{2}\nabla^{2}f_{3}-\Omega f_{3}-h_{2}\sin\alpha f_{1}-ih_{2}\cos\alpha f_{0}=0,\\
\xi_{F}^{2}\nabla^{2}f_{1}-\Omega f_{1}+h_{2}\sin\alpha f_{3}=0.\end{gathered}
\label{1}\end{equation}
 Here index $i=0,1,3$ stands for triplet condensate functions with
$0$ and $\pm1$ spin projections and for singlet condensate function,
$\xi_{F}^{2}=(D_{F}/2\pi T_{c})$, $D_{F}$ is diffusion coefficient
of F material, $\Omega=\omega/(\pi T_{c})$ are Matsubara frequencies,
and $h_{2}=H_{2}/(\pi T_{c})$. The system of Usadel equations for
the condensate functions $p_{i}$ for lower F film has the same form
as Eq. (1) with $\alpha=0$ and $h_{1}=H_{1}/(\pi T_{c})$ stands
instead of $h_{2}$.

Usadel equations must be supplemented by the boundary conditions.
At FF interface interface ($x=0$) they have the form \cite{KL}

\begin{equation}
\gamma_{B}\xi_{F}\frac{\partial}{\partial x}f_{i}+f_{i}=p_{i},\ \frac{\partial}{\partial x}p_{i}=\frac{\partial}{\partial x}f_{i},\ i=0,1,3,\label{2}\end{equation}
 here we consider that there is arbitrary transparency of FF interface
which is described by suppression parameter $\gamma_{B}$ \cite{GKI}.
At SF interface ($x=d_{F}$) we have

\begin{equation}
\xi_{F}\frac{\partial}{\partial x}p_{3}=\frac{\Delta}{\gamma_{BS}\sqrt{\Omega^{2}+\Delta^{2}}},\quad\frac{\partial}{\partial x}p_{0,1}=0,\label{3}\end{equation}
 where suppression parameter $\gamma_{BS}$ \cite{GKI} describes
SF interface and $\Delta$ is magnitude of order parameter in S film
normalized on $\pi T_{c}$ . Large value of $\gamma_{BS}$ permits
to neglect the suppression of order parameter in S film and consider
$\Delta$ in (\ref{3}) as only a temperature dependent value.

For simplicity we consider the limit of thin F films ($d_{F}/\xi_{F}<<1$).
In this limiting case the Green's functions in the first approximation
on $d_{F}/\xi_{F}$ are independent on space coordinates constants
and they can be found in similar way as in \cite{Karminskaya1}:

\begin{equation}
\begin{gathered}p_{1}=-\Gamma\frac{h_{2}\gamma_{BN}\sin(\alpha)S}{u^{2}\left(h_{2}^{2}\gamma_{BN}+v\right)\left(h_{1}^{2}\gamma_{BN}+v\right)-S^{2}}\text{,}\\
p_{0}=-i\Gamma\gamma_{BN}\frac{h_{2}\cos(\alpha)S-h_{1}u^{2}(v+h{}_{2}^{2}\gamma_{BN})}{u^{2}\left(h_{2}^{2}\gamma_{BN}+v\right)\left(h_{1}^{2}\gamma_{BN}+v\right)-S^{2}},\\
p_{3}=\Gamma\frac{\gamma_{BN}uv(h{}_{2}^{2}\gamma_{BN}+v)}{u^{2}\left(h_{2}^{2}\gamma_{BN}+v\right)\left(h_{1}^{2}\gamma_{BN}+v\right)-S^{2}},\\
f_{0}=-i\Gamma\frac{\gamma_{BN}u(h_{2}\cos(\alpha)S-h_{1}(v+\gamma_{BN}h{}_{2}^{2}))}{u^{2}\left(h_{2}^{2}\gamma_{BN}+v\right)\left(h_{1}^{2}\gamma_{BN}+v\right)-S^{2}},\\
f_{3}=-\Gamma\frac{\gamma_{BN}vS}{u^{2}\left(h_{2}^{2}\gamma_{BN}+v\right)\left(h_{1}^{2}\gamma_{BN}+v\right)-S^{2}},\\
f_{1}=P_{1}u\end{gathered}
\label{eq:funcgb}\end{equation}
 where parameter $\Gamma=\frac{\Delta\xi_{F}}{\gamma_{BS}d_{F}\sqrt{\Omega^{2}+\Delta^{2}}}$,
$\gamma_{BN}=d_{F}\gamma_{B}/\xi_{F}$ describes transparency of FF
interface, $S=h_{1}h_{2}\gamma_{BN}\cos(\alpha)-\Omega(u+1)$, $u=\Omega\gamma_{BN}+1$,
$v=\Omega(u+1)$.

In the case of collinear orientation of magnetization vectors, long-range
triplet components $p_{1}$ and $f_{1}$ are zero and it is convenient
to deal with complex condensate functions $p_{+}=p_{3}+p_{0}$ and
$f_{+}=f_{3}+f_{0}$. In the Matsubara representation, singlet components
$p_{3},f_{3}$ are real and short-range triplet components $p_{0},f_{0}$
are purely imaginary quantities.

For antiparallel orientation of magnetization vectors in both ferromagnetic
films ($\alpha=\pi$), functions $p_{3},f_{3}$ have the same sign
for any value of transparency of the FF interface, due to compensation
of magnetizations in the F-layers. This property is clearly seen from
Eq.~\ref{eq:funcgb}.

For parallel orientation of magnetizations ($\alpha=0$) the situation
is more complex since parameter S in Eq.~\ref{eq:funcgb} in this
case can change the sign as a function of $\gamma_{BN}$. Fig. 2 shows
the dependencies of real parts of condensate functions for middle
ferromagnet $Re(p_{+})=p_{3}$ (solid line) and for upper ferromagnet
$Re(f_{+})=f_{3}$ (dashed line) on the parameter $\gamma_{BN}$ for
the case $h_{1}=h_{2}=h$. For high transparent interface $p_{3}=f_{3}$,
while with the increase of $\gamma_{BN}$ the real part of condensate
function in upper ferromagnet changes sign at $\gamma_{BN}=\frac{2\Omega}{h_{1}h_{2}-\Omega^{2}}$.
This fact is due to the behavior of phases of complex functions $p_{+}$
and $f_{+}$. In Fig. 3 the dependencies of these phases ($Arg(p_{+})$
and $Arg(f_{+})$) are shown \textit{vs} the exchange field $h$.

In the high transparency regime, $\gamma_{BN}=0$, proximity coupling
of the upper and middle ferromagnetic films is strong, and the phases
of functions $p_{+}$ and $f_{+}$ coincide. However, for nonzero
$\gamma_{BN}$, the films can become effectively separated at large
values of $h$. Namely, with increase of $h$ the imaginary parts
of the condensate functions $p_{0},f_{0}$ increase, and phase slips
at both interfaces are generated. In accordance with the result of
\cite{GKF2002} for a single SF bilayer, the phase slips reach $-\pi/2$
at large $h$. As a result, the phase of upper F film shifts by $-\pi$
with respect to S, while the phase of middle film is saturated on
$-\pi/2$ at large $h$. The point where phase of the upper F film
crosses the value $-\pi/2$ corresponds to the sign change of the
singlet component $f_{3}$ in this film.

\begin{figure}[h]
 \centerline{\includegraphics{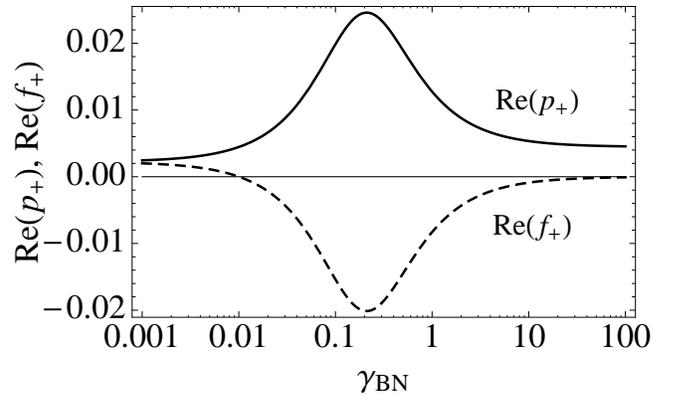}} \caption{$Re(p_{+})$ (solid line) and $Re(f_{+})$ (dashed line) \textit{vs}
parameter $\gamma_{BN}$ for misorientation angle $\alpha=0$, $\Omega=0.5$
and $h=10$.}

\end{figure}

\begin{figure}[h]
 \centerline{\includegraphics[scale=0.9]{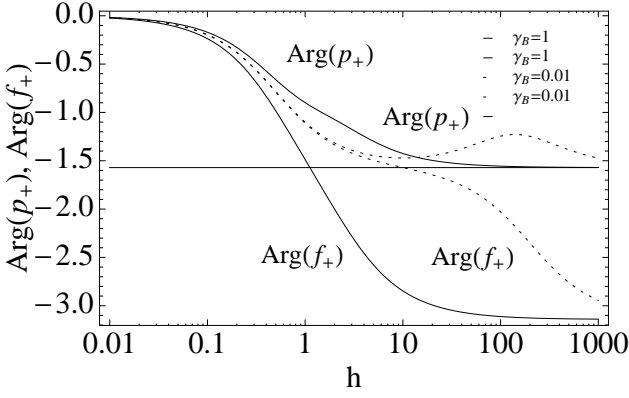}} \caption{$Arg(p_{+})$ and $Arg(f_{+})$ \textit{vs} value of exchange field
$h$ for misorientation angle $\alpha=0$, $\Omega=0.5$ and $\gamma_{BN}=1$
(solid line), $\gamma_{BN}=0.01$ (dashed line).}

\end{figure}

As a result, for sufficiently strong exchange field, singlet component
$f_{3}$ has opposite signs in parallel and antiparallel configurations.
Therefore, sign change of $f_{3}$ should occur at some intermediate
angle $\alpha$. The dependencies of condensate functions in upper
and middle ferromagnetic layers on angle $\alpha$ given by Eq.~(\ref{eq:funcgb})
are presented at Fig.4 for $h_{1}=10,h_{2}=30,\Omega=0.5$ for two
cases: $\gamma_{BN}=0$ and $\gamma_{BN}=0.01$.

\begin{figure}[h]
 \centerline{\includegraphics[scale=0.93]{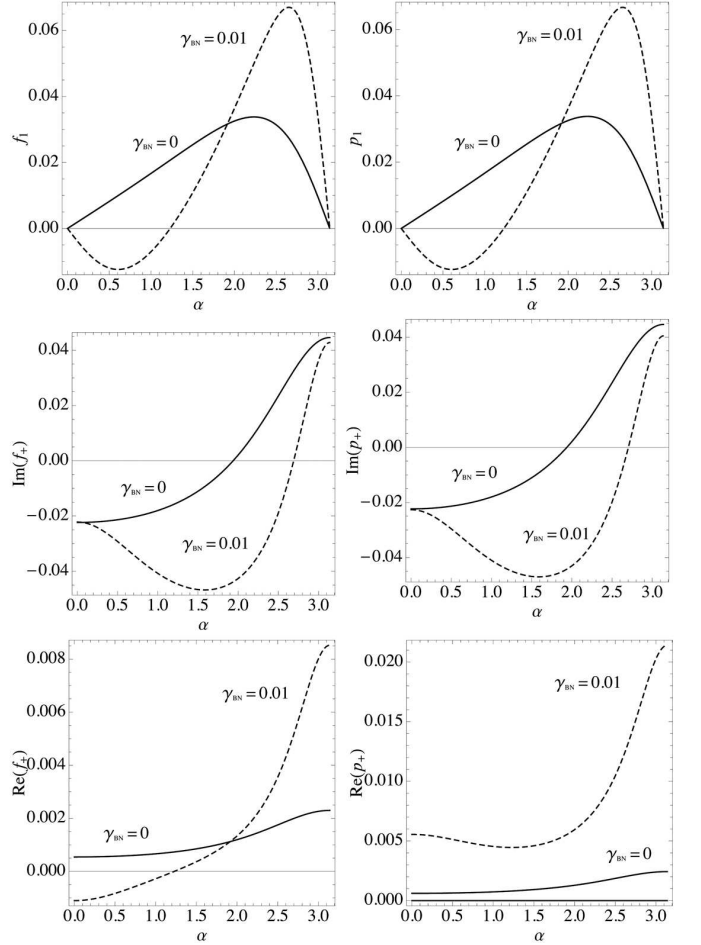}} \caption{Singlet components $Re(p_{+})$, $Re(f_{+})$ and triplet components
$Im(p_{+})$, $Im(f_{+})$, $p_{1}$, $f_{1}$ \textit{vs} misorientation
angle $\alpha$ at $h_{1}=10,\: h_{2}=30,\:\Omega=0.5$ and for $\gamma_{BN}=0,\:0.01$
(solid and dashed lines correspondingly).}

\end{figure}

It can be seen from Eq.~(\ref{eq:funcgb}) and Fig.4 that singlet
condensate function in the upper F film $f_{3}$ equals to zero at
an angle

\begin{equation}
\alpha_{in}=\pm\arccos(\frac{\Omega^{2}+\Omega/\gamma_{BN}}{h_{1}h_{2}}),\label{5}\end{equation}
 while the $p_{3}$ component in the lower F film has finite value.

This fact influences the behavior of triplet components $p_{1}$ and
$f_{1}$. If singlet component vanishes at least on one side of the
FF interface, the triplet components $p_{1}$ and $f_{1}$ will not
be generated in the system, while triplet components $p_{0}$ and
$f_{0}$ can still be nonzero in this case. As a result, triplet condensate
functions $p_{1},$ $f_{1}$ are zero not only at angles $\alpha=0,\:\pi$
but also at some intermediate angle $\alpha_{in}$ given by Eq.~(5),
if FF interface transparency has finite value.

It is clearly seen that the sign reversal effect is absent in limiting
case of transparent FF interface:

\[
\begin{gathered}f_{0}=p_{0}=i\Gamma\frac{h_{2}\cos(\alpha)+h_{1}}{h_{1}^{2}+h_{2}^{2}+2h_{1}h_{2}\cos(\alpha)+4\Omega^{2}}\\
f_{3}=p_{3}=\Gamma\frac{2\Omega}{h_{1}^{2}+h_{2}^{2}+2h_{1}h_{2}\cos(\alpha)+4\Omega^{2}}\\
f_{1}=p_{1}=\Gamma\frac{h_{2}\sin(\alpha)}{h_{1}^{2}+h_{2}^{2}+2h_{1}h_{2}\cos(\alpha)+4\Omega^{2}}\end{gathered}
\]

Another interesting effect is significant enhancement of the magnitudes
of singlet components for some range of values of parameter $\gamma_{BN}$
and related enhancement of triplet component with respect to transparent
FF interface.

These effects should lead to observable features in critical current
of SFFIS Josephson junction consisting from SFF trilayer coupled to
a superconductor S across tunnel barrier I. In this case, the current-phase
relation is sinusoidal with Josephson critical current given by simple
expression:

\[
I_{C}=\frac{\pi T}{eR_{I}}\sum_{n}f_{3}\frac{\Delta}{\sqrt{\Delta^{2}+\Omega^{2}}},\]
 where $R_{I}$ is the resistance of the interface I. The resulting dependence
of critical current on angle $\alpha$ is presented in Fig. 5 for
two different values of suppression parameter $\gamma_{BN}=0$ and
$\gamma_{BN}=0.1$.

\begin{figure}[h]
 \centerline{\includegraphics[scale=0.9]{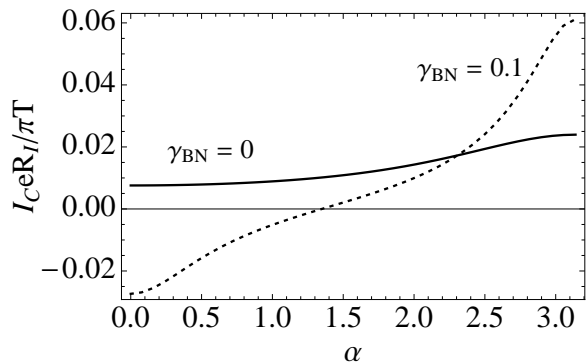}} \caption{$I_{C}$
 of SFFIS junction \textit{vs} misorientation angle $\alpha$, for $h_{1}=10,\: h_{2}=30,\: T=0.5T_{C}$,
$\gamma_{BN}=0$ (solid line) and $\gamma_{BN}=0.1$ (dotted line).}

\end{figure}

It is seen that critical current changes sign at some intermediate
angle for structure with nonzero $\gamma_{BN}$ on FF interface (dotted
line) in contrast with the case of zero $\gamma_{BN}$ (solid line).
The dependence of critical current on angle corresponds to dependence
of singlet condensate function $f_{3}$ (Fig. 4), however the angle
at which critical current changes sign does not coincide with the
one given by Eq.(5) because of summation over Matsubara frequencies
in the expression for critical current. To summarize, $0-\pi$ transition
may take place in SFFIS junction as function of misorientation angle
$\alpha$, if FF interface has finite transparency.

Interestingly, $0-\pi$ transition may also occur at zero $\alpha$
as a function of temperature, as shown in Fig.6. The low-temperature
critical current is very sensitive to the magnitude of $\gamma_{BN}$:
it is seen that critical current changes sign at low temperatures
in certain range of $\gamma_{BN}$ while at temperatures near $T_{C}$
critical current is still positive (solid line Fig. 6). Temperature-induced
$0-\pi$ transition was observed in Ref. \cite{ryazanov2001} in long
SFS junctions where $\pi$ state is realized due to oscillatory nature
of the condensate function in the F-layer. In the considered case
of SFFIS junction with thin ferromagnetic layers there are no oscillations
of the condensate functions in the F-layers, while the $\pi$ state
is realized due to accumulation of phase shifts at SF and FF interfaces.
With further increase of $\gamma_{BN}$ structure is in $\pi$ state
at all temperatures (dashed line in Fig.6).

\begin{figure}[H]
 \centerline{\includegraphics[scale=0.82]{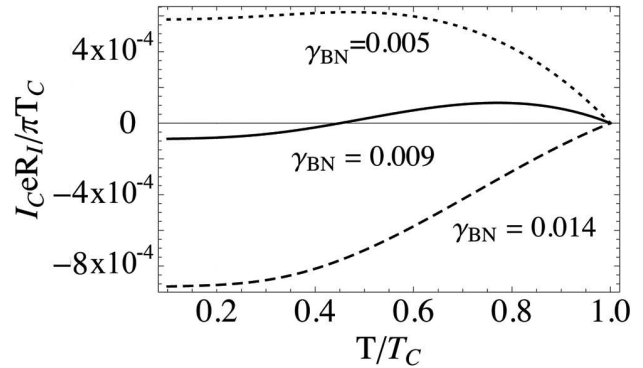}} \caption{$I_{C}$ of SFFIS junction \textit{vs} temperature $T$ for $\alpha=0$,
$h_{1}=10,\: h_{2}=30$, $\gamma_{BN}=0.005$ (dotted line) and $\gamma_{BN}=0.009$
(solid line), $\gamma_{BN}=0.014$ (dashed line).}

\end{figure}

In conclusion, we have investigated the proximity effect in SFF structures
with finite transparency of the FF interface. We have shown that due
to phase shift at the FF interface long-range triplet pair correlations
vanish not only at collinear orientations of magnetizations in both
layers, $\alpha=0,\pi$, but also at some intermediate angle $\alpha$.
This angle depends on parameters of the structure and typically is
close to $\pi/2$, when triplet correlations in symmetric FSF structure
are strongest. Moreover, maximum amplitudes of long-range triplet
and singlet pair correlations are achieved at finite transparency
of FF interface, not at ideal transparency, as can be expected. The
predicted effects manifest themselves in SFFIS Josephson junctions,
where the peculiarities of proximity effect in SFF trilayer lead to
possibility of realization of a $\pi$ state.

We acknowledge support from RFBR grant 11-02-12065-ofi-m.

\end{document}